\documentclass[journal,twocolumn]{IEEEtran}
\usepackage{xcolor}
\usepackage{mathtools}
\usepackage{epsfig,makeidx,color,subfigure}
\usepackage{amsmath,amssymb,bbm,enumitem}
\usepackage{cite,graphicx,lipsum}
\usepackage[switch,pagewise]{lineno}
\usepackage{tabularx}
\usepackage{hyperref}
\hypersetup{
        colorlinks = true,
        citecolor = red,
        breaklinks = true,
        linktocpage=true,
}
\usepackage{url}

\def\cK{{\cal K}}

\DeclareMathOperator*{\argmin}{\arg\!\min}
\DeclareMathOperator*{\argmax}{\arg\!\max}

\def\be{ \begin{equation} }
\def\ee{ \end{equation} }
\def\bea{ \begin{eqnarray} }
\def\eea{ \end{eqnarray} }

\def\bh{{\bf h}}

\def\b0{{\bf 0}}

\def\cD{{\cal D}}

\def\cN{{\cal N}}

\ifCLASSOPTIONonecolumn
\else

\fi

\begin{document}

\title{\fontsize{22}{28}\selectfont Reinforcement Learning for Opportunistic Routing in Software-Defined LEO–Terrestrial Systems}

\author{Sivaram Krishnan, Zhouyou Gu, Jihong Park, Sung-Min Oh, and Jinho Choi 
\thanks{
S. Krishnan and J. Choi are with the School of Electrical and Mechanical Engineering, University of Adelaide, Adelaide 5000, Australia (email: \{sivaram.krishnan, jinho.choi\}@adelaide.edu.au)

Z. Gu and J. Park are with the Singapore University of Technology and Design, Singapore (email: \{zhouyou\_gu, jihong\_park\}@sutd.edu.sg)

S. Oh is with the Electronics and Telecommunications Research Institute, Daejeon, Korea (email: smoh@etri.re.kr)

This study was supported by Institute of Information \& Communications Technology Planning \& evaluation (IITP) grant funded by the Korean government (MSIT) (No. RS-2024-00359235, Development of Ground Station Core Technology for Low Earth Orbit Cluster Satellite Communications).}
}

\maketitle
\begin{abstract}
The proliferation of large-scale low Earth orbit (LEO) satellite constellations is driving the need for intelligent routing strategies that can effectively deliver data to terrestrial networks under rapidly time-varying topologies and intermittent gateway visibility. Leveraging the global control capabilities of a geostationary (GEO)-resident software-defined networking (SDN) controller, we introduce \emph{opportunistic routing}, which aims to minimize delivery delay by forwarding packets to any currently available ground gateways rather than fixed destinations. This makes it a promising approach for achieving low-latency and robust data delivery in highly dynamic LEO networks. Specifically, we formulate a constrained stochastic optimization problem and employ a residual reinforcement learning framework to optimize opportunistic routing for reducing transmission delay. Simulation results over multiple days of orbital data demonstrate that our method achieves significant improvements in queue length reduction compared to classical backpressure and other well-known queueing algorithms.
\end{abstract}


\ifCLASSOPTIONonecolumn
\baselineskip 28pt
\fi

\section{Introduction}

Satellite communication systems have traditionally been developed independently from terrestrial networks, often with distinct design goals and architectures. Most early satellite systems were primarily intended for signal relay between ground stations and remote locations on Earth, typically using satellites as passive or transparent repeaters \cite{Evans2011satellite}. However, recent advances—especially the deployment of low Earth orbit (LEO) satellite constellations—have sparked increasing interest in enabling inter-satellite links (ISLs), which allow LEO satellites to communicate directly with one another \cite{Homssi22}. This shift, driven by regenerative payloads \cite{Yahia25}, enables autonomous LEO networks and their integration with terrestrial infrastructure \cite{Gopal18, Boero18}. For beyond-5G/6G NTNs, such integration promises global coverage and low latency, but also demands new architectures and control frameworks tailored to LEO dynamics \cite{ Saad24, Jamshed25}.


Software-defined networking (SDN) offers centralized, software-based control for satellite systems, enhancing flexibility and scalability over traditional distributed architectures \cite{Kreutz15}. It aligns well with LEO constellations, whose predictable motion allows accurate modeling of network dynamics and supports a global, time-evolving view in SDN. 
Studies show SDN improves handover performance \cite{Yang16} and enables efficient routing through global network awareness \cite{Kumar22}. 
SDN-based control is particularly beneficial in complex scenarios involving multiple LEO constellations operating at different altitudes or in distinct orbital planes. By centralizing control, potentially at the geostationary orbit (GEO) level, inter-satellite communication on ISLs can be systematically managed and optimized. 
Furthermore, this paradigm also supports the integration of LEO satellite systems as extensions of terrestrial infrastructures, i.e., enabling a LEO-terrestrial system.
In this context, backhaul transmissions from remote devices, e.g., sensors, do not require a fixed end-to-end path; instead, packets can be forwarded to any satellite connected to a ground gateway providing the terrestrial backbone. This flexible approach, referred to as \emph{opportunistic routing}, is the focus of our study, and differs from conventional static routing, which relies on designated source–destination pairs. Opportunistic routing holds promise for enabling low-latency applications such as industrial IoT telemetry.
\begin{figure}[!h]
    \centering
    \includegraphics[width=0.85\linewidth]{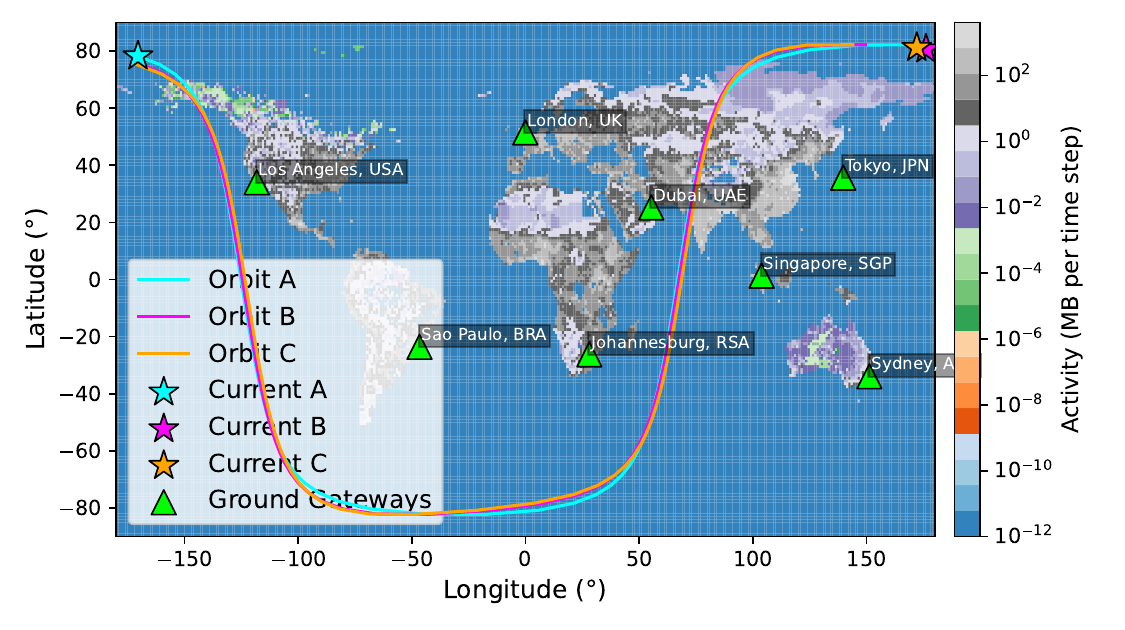}
    \caption{Aggregated activity rate over the globe with the ground-track of three Starlink satellites (Orbit A, Orbit B, Orbit C), together with gateway locations and their access periods. Aggregated activity denotes the total user traffic demand (MB per time step) summed over the satellite footprint.}
    \label{Fig:2d}
    \vspace{-0.25cm}
\end{figure}

\subsection{Our Contributions}
We consider a GEO satellite that acts as a central controller \cite{Kreutz15} coordinating the LEO satellite network providing connections to terrestrial infrastructures. 
Here, the GEO satellite operates the control plane and each LEO satellite functions as a switch of the data plane, enabling SDN in LEO-terrestrial systems.
In particular, as discussed in \cite{Yang16} \cite{Kumar22}, the GEO satellite provides control signals to the LEO satellites, while the LEO satellites send their status information back to the GEO satellite. Notably, during most of their orbital period, LEO satellites remain idle with respect to data collection and have only limited time windows for offloading buffered data to ground gateways. This constraint is illustrated in Fig.~\ref{Fig:2d}, which shows the Starlink orbits together with their access opportunities to well-known ground gateways. 
Leveraging the wide signaling coverage over the LEO constellation, the GEO-satellite controller can coordinate ISL switching intelligently to facilitate efficient packet forwarding toward satellites with gateway accesses.
The contributions of the paper are listed as follows.

\begin{figure}[t]
    \centering
    \includegraphics[width=0.8\linewidth]{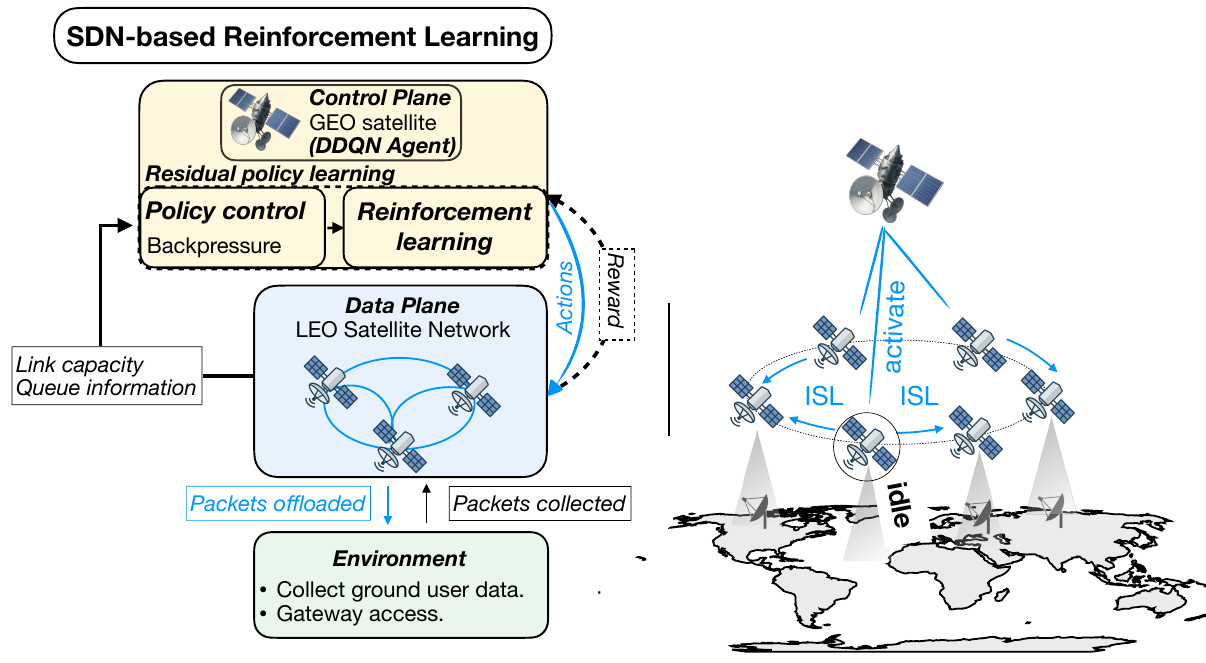}
    \caption{Residual reinforcement learning for opportunistic routing in the LEO-terrestrial system.}
    \label{Fig:prob}
    \vspace{-0.25cm}
\end{figure}

\begin{itemize}
\item We formulate SDN for opportunistic routing in LEO constellations with centralized GEO control, where packets are forwarded to any satellite with the available gateway to minimize delay. Unlike the architectures considered in \cite{Yang16, Kumar22}, where SDN-based LEO networks are studied in isolation from terrestrial networks, this paper assumes an integrated LEO-terrestrial system.
\item For our problem scenario, we incorporate \emph{spatio-temporal} traffic generation using population-density data and simulate actual satellite orbits to capture realistic queue evolution, ISL capacities, and gateway availability, achieving a realistic LEO-terrestrial system model.
\item We design a residual policy framework (Fig.~\ref{Fig:prob}) that augments a queue-stabilizing baseline such as backpressure with reinforcement learning. This retains a stabilizing backpressure bias while adapting to dynamic network conditions, and achieves superior performance over classical algorithms in both mean and peak queue metrics.
\end{itemize}

\section{System Model}
This section models a LEO satellite network that collects traffic from ground users and forwards it to terrestrial gateways through ISLs. The model captures the satellite constellation and its dynamic connectivity, the set of ground gateways acting as entry points to the terrestrial backbone, and the evolution of traffic queues at each satellite. 



\subsection{Network Structure}

Let $\cK = \{1, 2, \cdots, K\}$ denote the total number of LEO satellites in consideration. The position of the satellite $k \in \cK$ at time slot $t$ is   represented by its Earth-centered inertial (ECI) coordinate, denoted as $\mathbf{l}_k(t) = [x_k(t), y_k(t), z_k(t)] \in \mathbb{R}^3$. The time of the satellite network is synchronized and is slotted with the indices $t = 1, 2, \cdots, T$, with the slot duration being $\Delta t$ step long, in seconds. Since each satellite can only maintain a limited number of active inter-satellite links (due to finite number of transceivers), we define the neighborhood of satellite $k$ at time $t$ using 
\begin{align}
& \mathcal{N}_k(t)
= 
\argmin_{\mathcal{J} \subseteq \mathcal{K} \setminus \{k\},\ |\mathcal{J}| = M}
\;\sum_{j \in \mathcal{J}}
\mathcal{D}_{\mathrm{ECI}}\!\big(\mathbf{l}_k(t), \mathbf{l}_j(t)\big) \cr
& \text{subject to} \
\left\{
\begin{array}{l}
\mathcal{D}_{\mathrm{ECI}}\!\big(\mathbf{l}_k(t), \mathbf{l}_j(t)\big) \le R_{\max}, \ \forall j \in \mathcal{J}, \cr
\boldsymbol{\Phi}\!\big(\mathbf{l}_k(t), \mathbf{l}_j(t)\big) \le \Theta, \ \forall j \in \mathcal{J}, \cr
\end{array}
\right.
\label{eq:neighbor-selection}
\end{align}
where the inter-satellite Euclidean distance is
$\mathcal D_{\mathrm{ECI}}(\mathbf a(t),\mathbf b(t))=\|\mathbf a(t)-\mathbf b(t)\|_2$
(in km), with a range mask $R_{\max}$ (km). The plane-separation angle
$\boldsymbol{\Phi}(\mathbf l_k(t),\mathbf l_j(t))$ must satisfy
$\boldsymbol{\Phi}\le \Theta$ (radians). We also assume each satellite maintains at most $M\in\mathbb N$ neighbors.

We consider the presence of a set of $G$ gateways, denoted using the set representation $\mathcal{G} = \{1, 2, \cdots, G\}$. The location of gateway $g$ is represented in the ECI frame as $\bh_g(t) \in \mathbb{R}^3$, which is obtained by converting its geodetic latitude/longitude into the ECI coordinates at time $t$. These gateways serve as entry points to the terrestrial backbone network and act as final destinations for data transmitted through the satellite constellation. It is assumed that LEO satellites are capable of offloading buffered data packets to a gateway whenever it lies within the satellite’s communication footprint, given using 
\begin{align}
A_k(t) 
\!=\! \bigg\{\! \mathbf{h} \in \mathbb{R}^3 \big| \|\mathbf{h}\|_2 = R_E,
\frac{\big(\mathbf{l}_k(t) - \mathbf{h}\big)^\top \mathbf{h}}
{R_E \, \|\mathbf{l}_k(t) - \mathbf{h}\|_2}\ge \sin \varepsilon_{\min}\! \bigg\}.
\end{align}
where $R_E$ is the Earth's radius (in km) and $\varepsilon_{\min}$ is the minimum elevation angle (in radians) required for communication. 
At each time step $t$, satellite $k$ associates with the closest gateway within its coverage area $A_k(t)$
\be
g_k^\ast(t) 
= \underset{g \in \mathcal{G}:\ \mathbf{h}_g(t)\in A_k(t)}{\arg\min} 
\;\; \cD_{\text{ECI}}\!\left(\mathbf{l}_k(t), \mathbf{h}_g(t)\right)
\ee


\subsection{Queue Modeling}
\subsubsection{Packet Arrival Model}
Each LEO satellite in the LEO network is a node that receives packets from ground users within its footprint, and offloads stored packets either to an accessible ground gateway (if one lies within its coverage radius), or opportunistically to neighboring LEO satellites within $\cN_k(t)$ for further relaying.  Let $Q_k(t)$ represent the queue for the LEO satellite $k$. 

Over the coverage area $A_k(t)$, the aggregate arrival rate is
\be
\lambda_k(t)=M_{\text{tod}}\!\big(h(t)\big)\!\int_{A_k(t)} \rho_k(\mathbf r)\,\mathrm dA,
\ee
where $\mathbf r$ denotes a ground location, $M_\text{tod}(h)=\alpha\sin\!\big(2\pi(h-\tau)/24\big)+\beta$ with $h$ the hour of day, $\rho_k(\mathbf r)$ is the spatial traffic intensity. Additionally $\alpha\!\ge\!0$ is the diurnal amplitude, $\beta\!>\!0$ is the baseline offset, and $\tau$ is the peak-hour phase. The  arrivals are modeled as a Poisson random variable
\be
u_k(t)\sim\mathrm{Pois}(\lambda_k(t))
\ee

\subsubsection{Ground to channel link} 

It is also assumed that the transmitted packets to a ground gateway by LEO satellite $k$, $v_k (t)$, is a random variable, while its distribution is known. The ground gateway is usually equipped with a directional antenna. Thus, without severe interference, a reasonably high data rate can be achieved under good channel conditions. 
Let $\bar\Gamma_k (t)$ denote the signal-to-noise ratio (SNR) for the link from LEO $k \in \{1,\ldots, K\}$ to its associated closest ground gateway, which is given by
\begin{align}
\bar\Gamma_k(t) 
= 
\frac{P_{{\rm tx},k} \, 
\mathcal{D}_{\mathrm{ECI}}\!\big(\mathbf{l}_k(t), \mathbf{h}_{g_k^\ast}(t)\big)^{-\eta}}
{\sigma^2},
\end{align}
where $P_{\mathrm{tx},k}$ is the transmit power, $\eta$ is the path-loss exponent, and $\sigma^2$ is the noise variance. The short-term channel fading is modeled by a shadowed-Rician distribution \cite{Loo85}, with small-scale fading gain $\kappa(t)$ having pdf  
$f(\kappa) = \left( \frac{2 \xi_0 m_0}{2 \xi_0 m_0 + \Omega_0}
\right)^{m_0} \frac{e^{-\frac{\kappa}{2 \xi_0}}} {2 \xi_0} 
\mathstrut_1  F_1 \left(m_0, 1, \frac{\Omega_0 \kappa } {2 \xi_0 (2 \xi_0 m_0 + \Omega_0)} 
\right)$,
where $\xi_0$, $m_0$, and $\Omega_0$ are the functions of the elevation angle, denoted by $\psi$, and $\mathstrut_1  F_1$ is the confluent hypergeometric function. In \cite{Abdi03}, $\xi_0$, $m_0$, and $\Omega_0$ are determined as polynomial functions of $\psi$ through polynomial fitting. Taking $\kappa(t)$ as a small-scaling fading term, we can show that $\Gamma_k (t) = \kappa(t) \bar\Gamma_k (t) $. Let $L_{\text{pkt}}$ denote the (fixed) packet size in bits. Then, the achievable rate (in terms of the spectral efficiency) becomes
$R_k (t) = \log_2 (1+ \Gamma_k (t))$. 
In particular, we can see that 
\begin{equation}\label{EQ:d_k}
    \begin{aligned}
        v_k(t)\!\! =\!\! 
        \begin{cases}
            \min{(Q_k(t), D^{\text{LG}} R_k(t))}, \text{if a gateway is available,} \\
            0, \text{otherwise,}
        \end{cases}
    \end{aligned}
\end{equation}
where $D^{\text{LG}} = \frac{B\,\Delta t}{L_{\text{pkt}}}$
 is a constant that depends on the system bandwidth ($B$), the length of time slot, and so on. 

\subsubsection{ISL links}: 
Similarly, the large-scale SNR for the ISL from satellite $k$ to a neighbor 
$m \in \mathcal{N}_k(t)$ follows a free-space path loss model as
\begin{align}
{\Gamma}_{k \to m}(t) 
= 
\frac{
P_{\mathrm{tx},k} \, 
G_{\mathrm{tx},k} \, 
G_{\mathrm{rx},m} \,
\bigl(\frac{\lambda_c}{4 \pi \mathcal{D}_{\mathrm{ECI}}(\mathbf{l}_k(t), \mathbf{l}_m(t))}\bigr)^2
}{
k_B T_s B^{\mathrm{ISL}}
},
\end{align}
where $G_{\text{tx}, k}$ and $G_{\text{rx}, m}$ are the transmit and receive antenna gains, respectively. $k_B$, $T_s$, $\lambda_c$ and $B^\text{ISL}$ represent the Boltzmann's constant, system noise temperature, carrier wavelength and ISL channel bandwidth, respectively. 
We denote $w_{k \to m}(t)$, which represents the number of packets that the satellite $k$ forwards to its neighbor $m \in \cN_k(t)$. This decision is constrained by the capacity of the specific ISL channel, such that:
\be
0 \le w_{k \to m} (t) \le D^{\text{ISL}} \log_2 (1 + \Gamma_{k \to m} (t)), \label{eq:uplim} 
\ee
where $D^{\text{ISL}} = \frac{B^{\text{ISL}}\,\Delta t}{L_{\text{pkt}}}$
is a constant depending on the bandwidth ($B^\text{ISL}$) allocated to the inter-satellite links.

Let  $v_k(t)$ represent the total number of packets transmitted from LEO satellite $k$ to its nearest ground gateway. While the total number of transmitted and received packets over ISLs for satellite $k$ is given using
\be
d_k(t) = \sum_{m \in \mathcal{N}_k(t)} w_{k \to m}(t), \quad r_k(t) = \sum_{i \in \mathcal{N}_k(t)} w_{i \to k}(t),
\ee
and the total transmitted packets must satisfy the queue feasibility constraint 
\be
v_k(t) + d_k(t) \le Q_k(t). \label{eq:queuea}
\ee

\subsubsection{Queue Dynamics} 
The queue state $Q_k(t)$ of satellite $k$ evolves over time based on packet arrivals and transmissions:
\begin{equation}
Q_k(t+1) = \left( Q_k(t) + u_k(t) + r_k(t) - v_k(t) - d_k(t) \right)^+, \forall k \label{eq:quedy}
\end{equation}
where $(x)^+ = \max(0, x)$.

\subsection{Optimization Problem Formulation}
Given the dynamics in \eqref{eq:quedy}, the overall goal is to design a control strategy that decides the forwarding actions $\{w_{k \to m}(t)\}$ such that the network remains stable while efficiently utilizing both ground gateways and the other ISLs. The objective is to minimize queue congestion by minimizing the maximum queue length as
\begin{align}
\min_{w_{k\to m}(t)} \mathbb{E}\left[\max_k Q_k (t+1)\right] \ 
 \text{subject to} \
\eqref{eq:uplim},\eqref{eq:queuea},\eqref{eq:quedy}
\label{eq:opt}
\end{align}

\section{SDN-based Reinforcement Learning Approach}
Traditional scheduling algorithms like MaxWeight \cite{tassiulas1990stability} and backpressure \cite{tassiulas1990stability} are proven to maintain network stability and achieve maximum throughput. 
However, applying them directly is challenging due to the rapid change of neighborhoods, non-stationary arrivals, and time-varying channels.
Moreover, due to the geographical distribution of the LEO satellites and the terrestrial infrastructure, each satellite only has access to local information on the constellation, making it challenging for satellites to collaborate across the constellations to make optimal scheduling decisions.

\subsection{Proposed Solution}
To further enhance adaptability under dynamic traffic and topology changes \emph{reinforcement learning}\cite{Sutton19} provides a natural alternative by framing the routing problem as a sequential decision process. The agent can, in principle, adapt its policy $\pi$ to both the stochastic traffic arrivals $u_k(t)$ and the non-stationary network topology $\cN_k(t)$, with the objective of minimizing long-term congestion as described in \eqref{eq:opt}. Crucially, although the network topology changes rapidly, it exhibits quasi-periodic patterns driven by orbital mechanics, allowing the reinforcement learning agent to exploit these patterns and improve over purely reactive scheduling rules. 

Furthermore, enabling the SDN architecture to incorporate the reinforcement learning framework leverages the strength of \emph{global coordination} over the GEO satellite and the \emph{dynamic learning policy} via training neural networks as the policies. The SDN-based reinforcement learning framework can benefit from the global view of the network provided by the SDN controller on the GEO satellites, enabling centralized decision-making, while still allowing for distributed packet forwarding in the data plane.

\textbf{Challenges:} Nevertheless, directly applying reinforcement learning can be challenging as follows:  1) The action space is not fixed but varies with the number of neighbors $|\cN_k(t)|$, as seen in \eqref{eq:neighbor-selection}. Furthermore, the dimensionality of the action vector $\{w_{k \to m}(t)\}$ grows with the constellation size $K$ and neighborhood size $M$, making naive exploration intractable and leading to poor performance. 2) The reward feedback is delayed and highly coupled, since the queue dynamics in \eqref{eq:quedy} depend jointly on all agents' actions across the network.

To address these issues, we adopt a residual policy learning framework \cite{silver2018residual}, which augments a strong baseline controller with a learned residual term. Specifically, we use a backpressure-based policy \cite{tassiulas1990stability} as the baseline. As depicted in Fig.~\ref{Fig:prob}, the learned policy then outputs an additive correction (or residual) to the backpressure-derived action, enabling the agent to adapt to network dynamics while preserving desirable baseline properties. The backpressure policy determines link activation based on the score
\be
\pi_{\text{BP}}:s_{k \to m}^{\mathrm{BP}}(t) 
= \bigl(Q_k(t) - Q_m(t)\bigr) C_{k \to m}^{\mathrm{ISL}},
\ee
where $C_{k \to m}^{\mathrm{ISL}} = D^{\mathrm{ISL}}\log_2(1+\Gamma_{k \to m}(t))$ is the ISL capacity and a link is activated between two satellites $k$ and its neighbor $m$ if $s^\text{BP}_{k \to m } > 0$. Although backpressure guarantees stability and throughput optimality, it does not directly account for time-varying gateway availability or the rapidly changing network 
topology in LEO constellations. 

\subsection{Proposed Approach}  

To address these limitations, we introduce our residual reinforcement learning. We define the system state at time $t$ as
\begin{equation}
S(t) = \{ Q_k(t), v_k(t), d_k(t), r_k(t) : k \in \cK \} .
\end{equation}
The binary action set is given by
\begin{equation}
a(t) = \{a_{k \to m} : m \in \cN_k(t), k \in \cK\}, \quad 
a_{k \to m}\in\{0,1\},
\end{equation}
with scheduled packets given as: $w_{k \to m}(t) = a_{k \to m}(t)\, D^{\mathrm{ISL}} \log_2 \!\bigl(1+\Gamma_{k \to m}(t)\bigr)$.

We denote our residual policy as $\pi_{\mathrm{LG-BP}}$, which augments the original backpressure policy $\pi_{\mathrm{BP}}$ by incorporating link-to-ground (LG) awareness. The link-activation scores are
\begin{align}
\pi_{\text{LG-BP}}: s_{k \to m}^{\mathrm{LG}}(t) 
&= s_{k \to m}^{\mathrm{BP}}(t) 
+ \lambda^{\mathrm{LG-BP}} C_{m}^{\mathrm{LG}}, 
\label{eq:lg-score}
\end{align}
where $C_{m}^{\mathrm{LG}} = D^{\mathrm{LG}}\log_2(1+\Gamma_{m}^{\mathrm{LG}}(t))$ 
is the downlink capacity of neighbor $m$ and $\lambda^{\mathrm{LG-BP}}$ is a weight balancing ISL backlog reduction against downstream LG availability.
The reward function is given as:
\begin{align}
R(t) & = - \Big[
\alpha  \left(\bar{Q}_{\rm a}(t{+}1)-\bar{Q}_{\rm BP}(t{+}1)\right) \notag\\
&\quad + \beta  \left(Q^{\max}_{\rm a}(t{+}1)-Q^{\max}_{\rm BP}(t{+}1)\right)
\Big] ,
\label{eq:rrl-reward}
\end{align}
where $\bar{Q}_{\rm a}(t+1)$ and $Q^{\max}_{\rm a}(t+1)$ are the mean and maximum queue lengths under the agent’s actions, respectively, and $\bar{Q}_{\rm BP}(t+1)$ and $Q^{\max}_{\rm BP}(t+1)$ are the corresponding values under backpressure, respectively. The weights $\alpha,\beta > 0$ control the trade-off between reducing overall congestion and protecting the worst-case queues. We learn a policy $\pi$ by maximizing the expected discounted return  as follows:
\begin{equation}
\pi^\ast = 
\argmax_{\pi}\;\mathbb{E}_{\pi}\!\left[\sum_{t=0}^{\infty}\gamma^{t} R(t)\right],
\label{eq:disc-return}
\end{equation}
where $\gamma \in (0,1)$ represents the 
discount rate.

\section{Simulation Results}

We first evaluate our proposed approaches for opportunistic routing, namely Agent (the residual policy control), LG-BP, and vanilla reinforcement learning approach. Based on these results, we select the best-performing approach and compare it against the following baselines: {\bf Backpressure} \cite{tassiulas1990stability}, {\bf Equalize} (in which the most congested satellite forwards packets to its least loaded feasible neighbor at each step), {\bf No-ISL}, {\bf Max-Weight} \cite{tassiulas1990stability}, and {\bf Random}. 

\subsection{Environment Setup}
We use Space-Track \cite{spacetrack2025} APIs to obtain TLE data for satellite positions, and the 2020 Gridded Population of the World dataset to model ground traffic. The locations of the gateways are depicted in Fig.~\ref{Fig:2d}. 
Simulations span three consecutive days with slot length $\Delta t = 60$ s. Unless specified otherwise, the Starlink constellation with $K = 10$ satellites is used for the simulation, with $M=4$ neighbors per satellite. Queues for the satellites are assumed unbounded, so buffer overflow is not explicitly modeled, though reducing queue length inherently lowers overflow risk. To ensure robustness, the reported results are averaged over five independent runs of the evaluated methods. The environment is implemented in PyTorch and Gymnasium, and the reinforcement learning model was trained using the NVIDIA A100 GPU. 
\begin{figure}[!t]
    \centering    \includegraphics[width=0.7\linewidth]{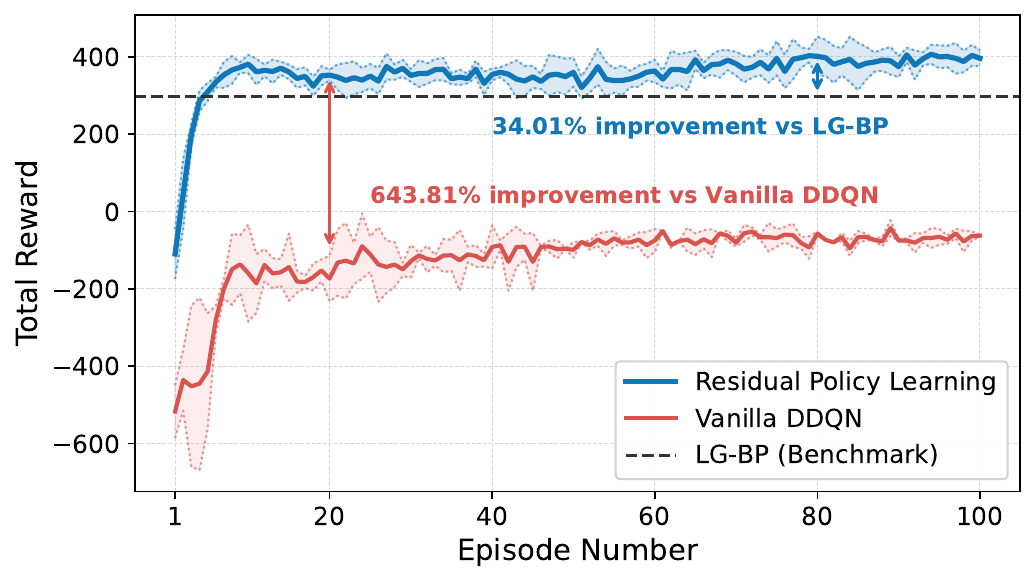}
    \caption{Training reward versus episode. The proposed residual-learning policy (blue) consistently outperforms the vanilla DDQN baseline (green).}
    \label{Fig:r0}
    \vspace{-0.25cm}
\end{figure}

\subsection{Results}
\subsubsection{Training Convergence}  To learn a residual policy over the backpressure baseline, we adopt a Double Deep Q-Network (DDQN) agent \cite{van2016deep}, which extends the classical Q-learning framework by approximating the state–action value function with a DNN, 
and by decoupling the action selection and action evaluation through a target network.

In Fig.~\ref{Fig:r0}, we compare our residual-learning DDQN with a vanilla DDQN baseline over 100 episodes under identical settings (3 layers, 256 hidden units each). Leveraging the LG-aware backpressure prior yields substantially higher rewards from the early episodes—achieving a 643.81\% improvement over vanilla DDQN and a 34.01\% improvement over the residual policy baseline. 

\subsubsection{Performance Against Other Methods} 

\begin{figure}[!t]
    \centering    \includegraphics[width=.85\linewidth]{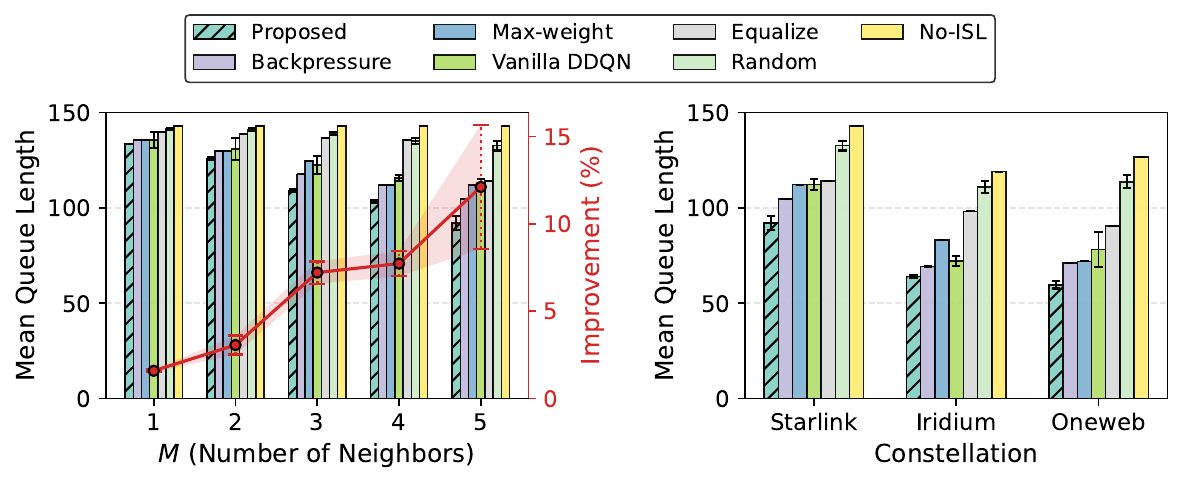}
    \caption{Queue length across varying maximum number of neighbors per satellite (left) and varying constellations (right) for different policies. }
    \label{Fig:r2}
    \vspace{-0.25cm}
\end{figure}

\begin{figure}[!t]
    \centering    \includegraphics[width=.85\linewidth]{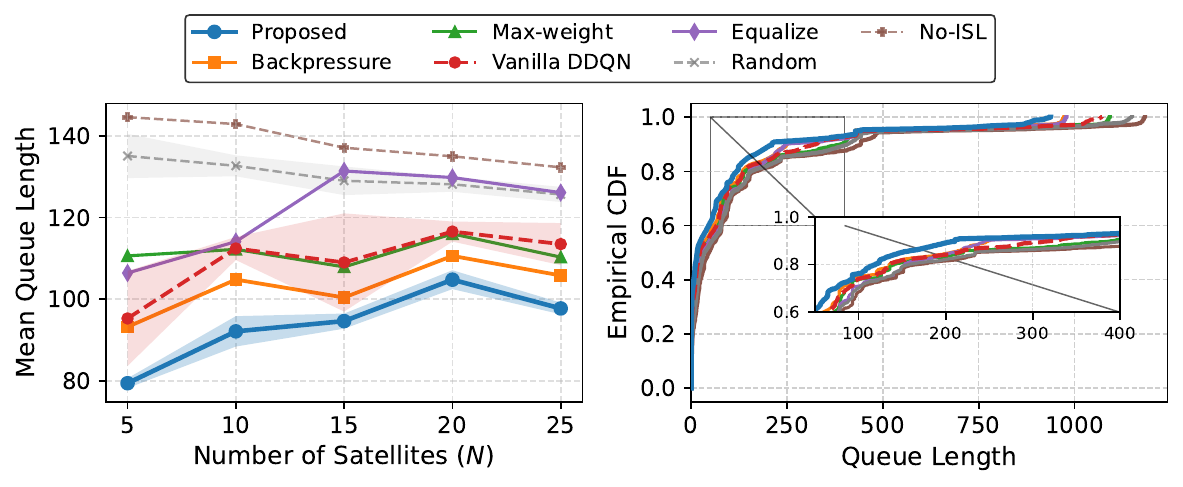}
    \caption{Performance evaluation: (a) Varying number of satellites, and (b) empirical CDF when $K = 10$ satellites.}
    \label{Fig:r1}
    \vspace{-0.25cm}
\end{figure}

Firstly, in Fig.~\ref{Fig:r2} (left), it can be seen that our approach outperforms the vanilla DDQN baseline, reducing mean queue length by 3.9–18.1\% across neighborhood sizes $M\in \{1,\ldots,5\}$, with varying maximum neighbors of each satellite.
Compared to classical policies, our method consistently yields the lowest queue lengths, with gains over backpressure increasing from 1.6\% at $M=1$ to 12.1\% at $M = 5$. In Fig.~\ref{Fig:r2} (right), we compare performance of various methods across Starlink, Iridium, and OneWeb constellations. Our method consistently lowers queue lengths, achieving an average reduction of 7.6–16.1\% compared to backpressure. In Fig.~\ref{Fig:r1} (left), we compare the mean queue length for different numbers of satellites in the constellation. Compared to the next-best performing approach, our residual method reduces congestion by 5.3 - 14.8\%. In Fig.~\ref{Fig:r1} (right), we also present the empirical cumulative distribution function (ECDF) of queue lengths observed over a single orbital period, which shows that our method outperforms.

\begin{figure}[!t]
    \centering    \includegraphics[width=0.7\linewidth]{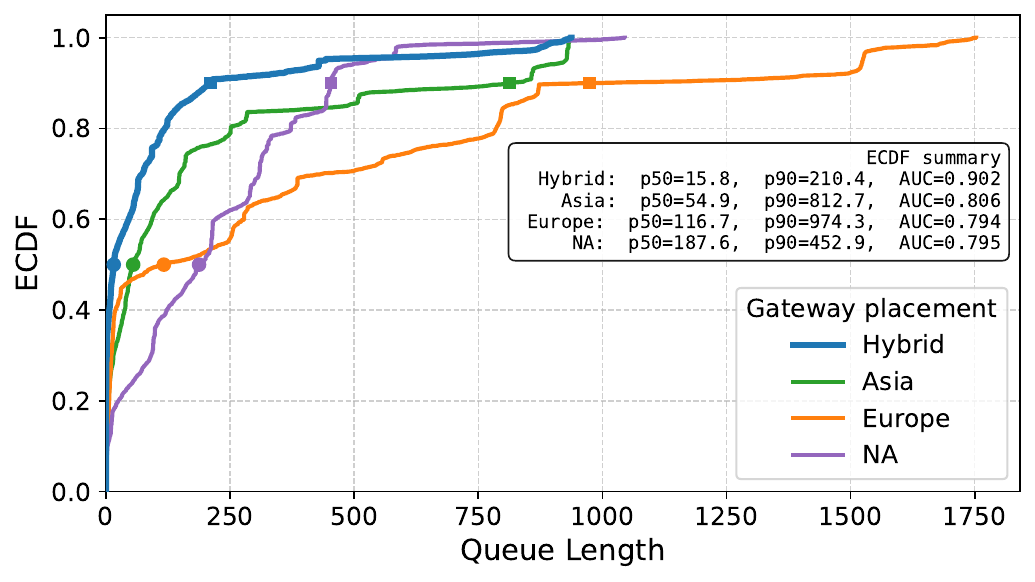}
    \caption{Empirical ECDF over varying placement of gateways.}
    \label{Fig:r5}
    \vspace{-0.25cm}
\end{figure}
\subsubsection{Varying Gateway Availability} We evaluate gateway placement in Fig.~\ref{Fig:r5}. The hybrid (global) deployment performs best by balancing load across regions. Among single-region cases, Asia performs relatively well, while Europe and North America suffer from limited coverage and bottlenecks.
\section{Conclusion}
We presented the SDN-based residual reinforcement learning approach for opportunistic routing in LEO satellite networks, leveraging a link-to-ground-aware backpressure baseline. Our method consistently reduces mean and peak queue lengths compared to the baselines, while remaining scalable over varying settings. Reinforcement learning shows particular promise for low-latency applications, as its adaptive decision-making can react to rapid topology changes and transient congestion, enabling faster delivery than fixed-rule schedulers. Future work will address scalability to mega-constellations, account for delays between the GEO controller and LEO satellites, and ensure consistent implementation of the reinforcement learning framework.
\bibliographystyle{ieeetr}
\bibliography{LEO_abbr}

\end{document}